# Measuring electron orbital magnetic moments in carbon nanotubes

E. D. Minot*, Yuval Yaish*, Vera Sazonova and Paul L. McEuen

*Laboratory of Atomic and Solid-State Physics, Cornell University, Ithaca, NY 14853*

*These authors contributed equally to the work

**The remarkable transport properties of carbon nanotubes (NTs) are determined by their unique electronic structure[1]. The electronic states of a NT form one-dimensional electron and hole subbands which, in general, are separated by an energy gap[2,3]. States near the energy gap are predicted to have a large orbital magnetic moment μ_orb much larger than the Bohr magneton[4,5]. The moment is due to electron motion around the NT circumference. This orbital magnetic moment is thought to play a role in the magnetic susceptibility of NTs[6-9] and the magneto-resistance observed in large multi-walled NTs[10-12]. However, the coupling between magnetic field and the electronic states of an individual NT has not been experimentally quantified. We have made electrical measurements of relatively small diameter (2 – 5 nm) individual NTs in the presence of an axial magnetic field. We observe energy shifts of electronic states and the associated changes in subband structure. Our results quantitatively confirm predicted values for μ_orb.**

The electronic structure of a NT is elegantly described by the quantization of wave states around a graphene cylinder[1]. Graphene is a zero band-gap semiconductor in which the valence and conduction states meet at two points in k-space, $\mathbf{K}_1$ and $\mathbf{K}_2$ (Fig. 1a). The dispersion around each of these points is a cone (Fig 1b). When graphene is wrapped into a cylinder the electron wave number perpendicular to the NT axis, $k_\perp$, is quantized, satisfying the boundary condition $\pi D k_\perp = 2\pi j$ where $D$ is the NT diameter and $j$ is an integer. The resulting allowed $\mathbf{k}$'s correspond to the horizontal lines in Fig.

1a that miss $\mathbf{K}_i$ by an amount $\Delta k_\perp$. The conic sections of the dispersion cones by allowed $\mathbf{k}$ determine the NT band structure near the Fermi level as shown in Fig. 1b. The upper and lower branches of the conic sections correspond to the conduction and valence states of the NT. Both the $\mathbf{K}_1$ and $\mathbf{K}_2$ subbands have the same energy gap between conduction and valence states: $E_g^0 = \hbar v_F \Delta k_\perp$.

The size of $\Delta k_\perp$, and therefore $E_g^0$, depends on the NT chirality[1-3] and perturbations such as curvature[13], axial strain[14,15], twist[14] and inner-outer shell interactions[16]. From consideration of chirality alone, NTs are classified as metallic ($\Delta k_\perp = 0$) or semiconducting ($\Delta k_\perp = 2/3D$)[1]. Perturbations displace the dispersion cones[13], modifying $\Delta k_\perp$ and resulting in an important class of small band-gap "quasi-metallic" NTs[17]. We have used these small band-gap NTs in our measurements.

The electron states near the gap correspond to semi-classical electron orbits encircling the NT. The perpendicular component of orbital velocity $v_\perp = (1/\hbar)\, dE/d k_\perp$ determines the clockwise (CW) or counterclockwise (CCW) sense of an orbit. For example, in Fig. 1b we see that $v_\perp$ is negative for the $\mathbf{K}_1$ conduction states but is positive for $\mathbf{K}_1$ valence states. By symmetry, each CW (CCW) orbit in the $\mathbf{K}_1$ subband has an equal energy CCW (CW) partner in the $\mathbf{K}_2$ subband. As a consequence, the two subbands are degenerate, but the CW/CCW sense of valence and conduction states is reversed.

From basic electromagnetic theory, an electron moving at velocity $v$ around a loop of diameter $D$ has an orbital magnetic moment of magnitude $\mu = Dev/4$. In a NT, electron states at the band-gap edges, where $v_\perp$ is largest, have an orbital magnetic moment of magnitude $\mu_{orb} = Dev_F/4$ directed along the tube axis. A magnetic field parallel to the NT axis, $B_\parallel$, is predicted to shift the energy of these states by:

$$\Delta E = -\mathbf{\mu}_{orb} \cdot \mathbf{B} = \pm \frac{Dev_F B_{\parallel}}{4}. \tag{1}$$

For NTs with a finite energy gap at $B_{\parallel} = 0$, the energy gap of one subband becomes larger as $B_{\parallel}$ is increased, while the energy gap of the other subband becomes smaller (Fig. 1c).

Previous work on the magnetoresistance of individual MWNTs[10-12] and the magnetic susceptibility of NT mats[6,7] has not confirmed the magnitude of $\mu_{orb}$ or the splitting of subband degeneracy. In the current work we use two different techniques to achieve this goal: (1) thermally activated transport through individual small band-gap NTs that are depleted of charge carriers, (2) energy level spectroscopy near the band-gap edge of NT quantum dots.

We have found that a suspended NT device geometry (Fig. 2) is well suited for studying small changes in band-gap. Measurements of many such devices, using a gold-coated atomic force microscope (AFM) tip as a movable, local electrode[18,19], show that NT segments contacting the oxide substrate are doped p-type, while suspended sections of the same tube are almost intrinsic. At small gate voltage $V_g$ the suspended section is depleted of charge carriers. The oxide bound sections, however, remain p-doped and act as electrodes to the suspended section. By studying the conductance of the suspended section at different temperatures and magnetic fields we can determine changes in $E_g^{K_i}$.

Figure 3a shows device conductance $G$ vs. $V_g$ of two small band-gap NTs. Device 1 shows a sharp dip near $V_g = 0.4$ V, corresponding to depletion of carriers in the suspended segment. A second, broader dip occurs at $V_g \approx 2$ V as the oxide-bound segments become depleted. The inset shows the dip from the suspended section of Device 2. In both cases, the addition of a magnetic field substantially increases the conductance at the bottom of the dip.

When the suspended NT segment is depleted, conductance occurs via thermal activation of carriers across the energy gap. Conductance is smallest at $V_g = V^*$, immediately before the suspended segment becomes n-type (Fig. 2c). The minimum conductance due to thermal activation, $G_{act}(V^*)$, can be estimated by considering the Fermi-Dirac function at temperature $T$ and the Landauer formalism for 1D conduction channels[15,20]

$$G_{act}(V^*,T) = \frac{2e^2}{h} \sum_{i=1,2} |t_i|^2 \frac{2}{\exp(E_g^{K_i}/k_B T)+1} \,, \qquad (2)$$

where $|t_i|^2$ is the transmission probability for thermally activated carriers in the $i^{th}$ subband. The device conductance $G$ is a combination of $G_{act}$ in series with the conductance of the p-type sections of NT and the conductance of the metal-NT contacts, both of which are largely temperature independent.

We have measured $G$ vs. $V_g$ for Devices 1 and 2 at several temperatures. In Fig. 3b (open circles) we plot the change in resistance $\Delta R(T) = G(V^*, T)^{-1} - G(V_g \ll 0, T)^{-1}$ of Device 2 at $B = 0$ T. From the slope and intercept of the fitting exponential, and assuming subband degeneracy ($E_g^{K_i} = E_g^0$) we find: $E_g^0 = 40$ meV and $|t_1|^2 + |t_2|^2 = 1.6$. Because $|t_1|^2 + |t_2|^2$ is close to 2, we conclude that transport is nearly ballistic and that both the $\mathbf{K}_1$ and $\mathbf{K}_2$ subbands make comparable contributions to the device conductance. We find similar values of $E_g^0$ in both devices (see Table 1) even though the NT diameters are significantly different. This suggests that the band-gaps are not curvature related[13]. Further work is needed to identify the perturbations responsible for $E_g^0$.

Magnetic fields dramatically reduce $\Delta R$, as shown in Fig 3c. The temperature dependence of $\Delta R$ at $B = 10$ T is also shown for Device 2 (Fig. 3b, black triangles). If we fit this high-field temperature data with the same method used for zero-field data, we find $E_g^0 = 22$ meV and $|t_1|^2 + |t_2|^2 = 0.8$. The band-gap of at least one subband is significantly lowered by the magnetic field and we argue below that the apparent change in $|t_1|^2 + |t_2|^2$ is due to the increasing band-gap of the second subband.

The magnetic field dependence of $\Delta R$ can be quantitatively described by equal and opposite changes in $E_g^{K_1}$ and $E_g^{K_2}$ due to the coupling of $\mu_{orb}$ with $B_\parallel$. We have accurately fit our measurements of $\Delta R(B,T)$ using Eq. (2) and setting $E_g^{K_1} = E_g^0 - aB$ and $E_g^{K_2} = E_g^0 + aB$ (see the fit curves in Fig. 3c). The only fit parameter is $a$; $E_g^0$ and $|t_i|^2$ are found from the temperature dependence of $\Delta R$ at $B = 0$ T and setting $|t_1|^2 = |t_2|^2$.

The fitting results for Devices 1 and 2 are summarized in Table 1. In agreement with Eq. 1, the measured $\mu_{orb}$ scale with diameter and are an order of magnitude larger than previously measured spin magnetic moments in NTs[21,22]. Thermally activated transport (Eq. (2)), combined with the breaking of CW/CCW subband degeneracy, describes $\Delta R$ over a wide range of $T$ and $B$. At $B = 10$ T device conductance is almost entirely due to carriers which are thermally activated across the smaller band-gap. Transport occurs in a single subband, explaining why $|t_1|^2 + |t_2|^2$ decreases by a factor of 2 when subband degeneracy is incorrectly assumed at high field. Our measurements confirm theoretical predictions[4,5] for the sign and magnitude of orbital magnetic moments in NTs and show that an applied magnetic field can split the degeneracy of the $\mathbf{K}_1$ and $\mathbf{K}_2$ subbands.

Orbital magnetic moments should also influence the energy level spectra of NT quantum dots (NTQDs) in applied magnetic fields. In our device geometry a NTQD forms when $V_g > V^*$ and electrons are confined to conduction states of the suspended section by p-n tunnel barriers (Fig. 2c). Figure 4a shows the formation of a NTQD in Device 1 at $V_g > V^*$, $T = 1.5$ K. There is a large region of zero conductance as the Fermi level passes through the energy gap of the suspended section. At higher $V_g$ the Coulomb diamonds labeled 1, 2, 3 and 4 correspond to charge states of one, two, three and four electrons in the conduction band of the suspended segment.

In the Coulomb blockade model of quantum dots[23] the width of the $N$th diamond is proportional to a fixed electrostatic charging energy plus the energy difference between the quantum levels occupied by $N$th and ($N$+1)th electrons. The energies of the quantum levels in our NTQD can be estimated by considering electrons confined to a 1D potential well of length $L$. The confinement results in quantized $k_{\parallel}$ values which, combined with the dispersion relations $E_i(k_{\parallel})$, determine the energy levels of the dot. Near the band-gap edge $E_i(k_{\parallel})$ are parabolic, therefore, the energy levels of the first few conduction states should be:

$$\varepsilon(n, i, B_{\parallel}) = \frac{E_g^0}{2} + \frac{\hbar^2 \pi^2}{2 m_i^* L^2} n^2 \pm \mu_{orb} B_{\parallel}, \qquad (3)$$

where the quantum number $n$ is a positive integer, the effective mass $m_i^* = E_g^{K_i}(B_{\parallel})/2 v_F^2$, and + applies to CW orbitals while – applies to CCW orbitals. The first few level crossings predicted by Eq (3) are shown in Fig. 4b.

Figure 4c shows low-bias $G$-$V_g$'s of a Coulomb peak from Device 1 as $B$ is increased. The peak corresponds to the second electron added to the dot (the intersection of Coulomb diamonds 1 and 2). The peak shifts ~ 1.2 mV/T and doubles in conductance as $B$ reaches 3.6 T. Figure 4d shows the first 8 Coulomb peaks. Peaks positions generally move between 1.2 and 1.6 mV/T. The fifth and subsequent peaks show clear changes between positive and negative slopes. Peaks appear to be paired, each pair having a different zigzag pattern.

The main features of Fig. 4d are described by the NTQD model. Peaks with $d\varepsilon/dB > 0$ correspond to tunneling into a CW orbital, while peaks with $d\varepsilon/dB < 0$ correspond to tunneling into a CCW orbital. The measured value of $\mu_{orb} = |\,d\varepsilon/dB_\parallel\,| =$ $0.7 \pm 0.1$ meV/T is inferred as described in the caption of Fig. 4d, and agrees with the values in Table 1 for Device 1. Furthermore, the striking difference between the first four peaks and later peaks is in qualitative agreement with the modeled spectrum (Fig. 4b). The first pair of peaks (spin up and spin down, $n = 1$, CCW orbital) are not expected to undergo level crossings. The second pair (peaks 3 and 4 in Fig. 4d) may undergo a level crossing at low field, however, the resolution of our data is limited by thermal broadening; levels separated by less than $4k_BT \approx 0.5$ meV merge together. The third and fourth pairs clearly show the changes in slope that are expected when level crossings occur. We conclude that there are quantum levels near the band-gap edges with both positive and negative orbital magnetic moments whose magnitudes are consistent with theoretical predictions[4,5]. The Coulomb blockade model does not describe all the features in Fig. 4d. The detailed structure of this NTQD system may depend on effects such as exchange coupling[21,24], and will be the subject of future work.

Our measured values of $\mu_{orb}$ are 10 - 20 times larger than the Bohr magneton and the spin magnetic moment in NTs[21,22]. The reason is the large size of electron orbits encircling the NT compared to the radii of atomic orbitals. These large magnetic moments give researchers a powerful new tool to control the energy structure of NTs. For example, the tunnel transparency of p-n barriers can be tuned by using a magnetic field to modify the band-gap. This effect is seen in Fig. 4c: the conductance of the Coulomb peak increases as the tunnel barriers become more transparent. This will be useful, for example, to study Kondo physics in NTQDs[24,25] at different tunneling strengths. Researchers can also tune the energy levels of electrons in the 1D box formed by a NT. By applying large magnetic fields it is possible to investigate the properties of a NT in which only one subband is occupied. Conversely, by matching the energies of different subband states, the interactions between states arising from CW and CCW orbits can be explored.

**Acknowledgements** We thank Hande Ustunel, Tomas Arias, and Hongjie Dai for useful discussions. This work was supported by the NSF through the Cornell Center for Materials Research and the Center for Nanoscale Systems, and by the MARCO Focused Research Center on Materials, Structures, and Devices. Sample fabrication was performed at the Cornell node of the National Nanofabrication Users Network, funded by NSF.  One of us (E.D.M.) acknowledges support by an NSF Graduate Fellowship.

**Correspondence** and requests for materials should be addressed to P.L.M. (mceuen@ccmr.cornell.edu).

**Figure 1** Nanotube states near the band-gap and orbital magnetic moments. **a,** The valence and conduction states of graphene meet at $\mathbf{K}_1$ and $\mathbf{K}_2$. Horizontal lines show the quantized values of $k_\perp$ for the NT structure in Fig. 1c. The misalignment between horizontal lines and the K points is $\triangle k_\perp$. **b,** Graphene dispersion near the K-points is described by the cones $E_i(\mathbf{k}) = \pm\hbar v_F|\mathbf{k} - \mathbf{K}_i|$, with $v_F$ = 8 x $10^5$ m/s (ref. 1). Lines of allowed $\mathbf{k}$ intersect the two cones (blue and red curves). The conduction states near $\mathbf{K}_1$ (upper blue curve) have $dE/dk_\perp$ < 0. Electrons in these states move around the NT in a counterclockwise (CCW) fashion. The valence states near $\mathbf{K}_1$ (lower blue curve) have $dE/dk_\perp$ > 0 and are associated with clockwise (CW) electron motion. CCW (CW) orbits correspond to positive (negative) magnetic moments along the NT axis. The conic section near $\mathbf{K}_2$ lies on the opposite face of an identical dispersion cone. Therefore, $\mathbf{K}_2$ conduction (valence) states have CW (CCW) orbits. **c,** Top, perspective view of a NT in the presence of a magnetic field $B_\parallel$. Below, the dispersion relations $E_1(k_\parallel)$ and $E_2(k_\parallel)$ shown in blue and red respectively. The subbands are degenerate at $B_\parallel$ = 0. The magnetic field breaks this degeneracy.

**Figure 2** Device geometry and band bending. **a,** NTs are grown on Si/SiO$_x$ substrates by the chemical vapour deposition method[26]. Electrodes (5nm Cr, 50nm Au) are patterned by photolithography[27]. The central region of the NT is suspended over a trench defined by electron beam lithography and wet etching using 6:1 buffered HF acid. **b,** AFM image of the suspended section of NT and nearby oxide-bound sections of Device 1. The scale bar is 130 nm. The suspended section appears fuzzy because it is displaced by the AFM tip during imaging. From the image we find NT diameter $D$ = 2.6 nm, suspended length $L$ = 500 nm, and determine the misalignment angle $\phi$ between applied magnetic field and the NT axis. **c,** Band bending in the suspended NT segment and neighboring oxide-bound segments when $V_g = V^*$. The number of thermally

activated carriers is minimized and there is no n-type region to facilitate tunneling processes. The oxide-bound sections remain p-type at small $V_g$.

**Figure 3** Effect of magnetic field on device resistance. **a,** $I$-$V_g$ curves for Devices 1 and 2 at T = 100K. Curves taken at $B$ = 0 T have lower conductance than curves taken at $B$ = 10 T. **b,** $\Delta R$ as a function of 1/$T$ for Device 2. The data shown are for $B$ = 0 T (larger $\Delta R$) and $B$ = 10 T (smaller $\Delta R$). **c,** $\Delta R$ as a function of $B$ for Device 1 at $T$ = 78 K (upper curve) and Device 2 at $T$ = 90 K (lower curve).

**Figure 4** Energy levels of a nanotube quantum dot. **a,** Differential conductance $dI/dV_{sd}$ as a function of source-drain voltage $V_{sd}$ and $V_g$. Data is from Device 1 at $T$ = 1.5 K. Dark blue represents $dI/dV_{sd}$ = 0, dark red represents $dI/dV_{sd}$ = 0.2 $e^2/h$. In the white regions (top and bottom of the plot) current levels exceeded the measurement range. The first four Coulomb diamonds, corresponding to discrete charge states, are labeled 1-4. The gate coupling $\alpha$ is twice the ratio of Coulomb diamond width to Coulomb diamond height[23]. For this device $\alpha \approx 2.2$. **b,** Modelled energies of quantum levels from Eq (3), approximating $m_i^*$ as constant. The energy scale $\delta = \hbar^2 \pi^2 / 2m_i^* L^2$. For Device 1 we have $\delta \approx 0.25$ meV. Coloured lines represent expected zigzags in first 6 Coulomb peaks with red and blue representing CCW and CW states respectively. Arrows indicate spin degeneracy for each state. **c,** Conductance $I/V_{sd}$ as a function of magnetic field $B$ for the second Coulomb peak of Device 1, $\phi$ = 30°, $V_{sd}$ = 0.5 mV. Shifts in peak position $V_g^p(n,i)$ are related to energy shifts of quantum levels by: $dV_g^p(n,i)/dB = \alpha \cdot d\varepsilon(n,i)/dB$. **d,** Low-bias conductance $I/V_{sd}$ as a function of $V_g$ and $B$ showing first 8 Coulomb peaks of Device 1, $\phi$ = 30°. Dark blue represents $I/V_{sd}$ = 0; dark red represents $I/V_{sd}$ = 0.35 $e^2/h$. The colour scale for peak 1 is magnified by 100 times.

## Table 1 Summary of thermal activation results

| | $D$ (nm) | $E_\mathrm{g}^0$ (meV) | $\phi$ (°) | a (meV/T) | $\mu_\mathrm{orb}$ (meV/T) | |
| --- | --- | --- | --- | --- | --- | --- |
| | | | | | Experiment | Theory |
| Device 1 | 2.6 ± 0.3 | 36 ± 3 | 30 ± 3 | 1.3 ± 0.1 | 0.7 ± 0.1 | 0.5 ± 0.1 |
| | | | 60 ± 3 | 0.7 ± 0.1 | 0.7 ± 0.1 | 0.5 ± 0.1 |
| Device 2 | 5.0 ± 0.3 | 40 ± 3 | 45 ± 3 | 2.1 ± 0.2 | 1.5 ± 0.2 | 1.0 ± 0.2 |

$\phi$ is the misalignment angle between NT axis and the magnetic field direction. The experimental value of $\mu_\mathrm{orb}$ is given by $a/2\cos\phi$. There is uncertainty in theoretical values of $\mu_\mathrm{orb}$ due to uncertainty in $v_\mathrm{F}$ and $D$.

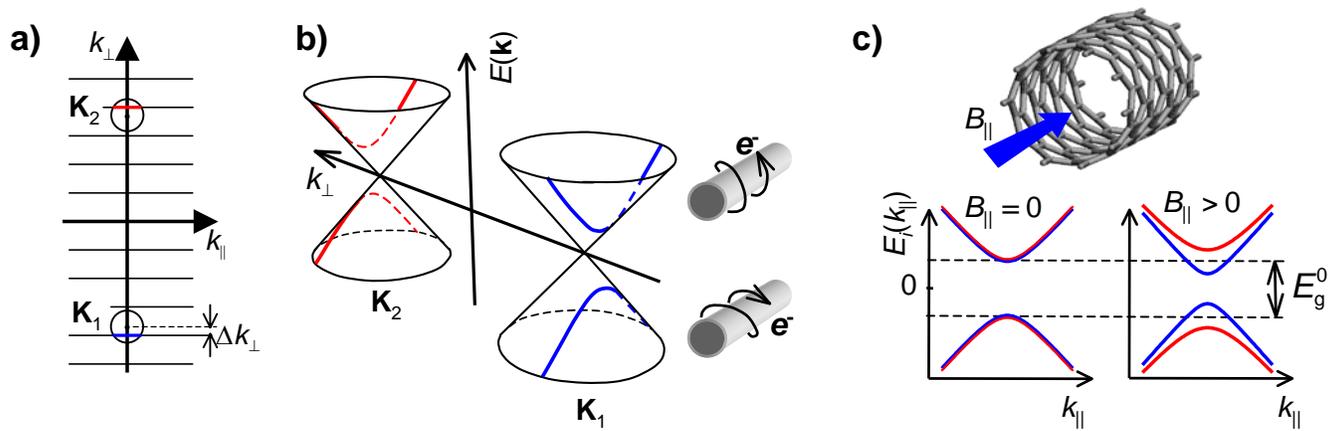

**Figure 1**

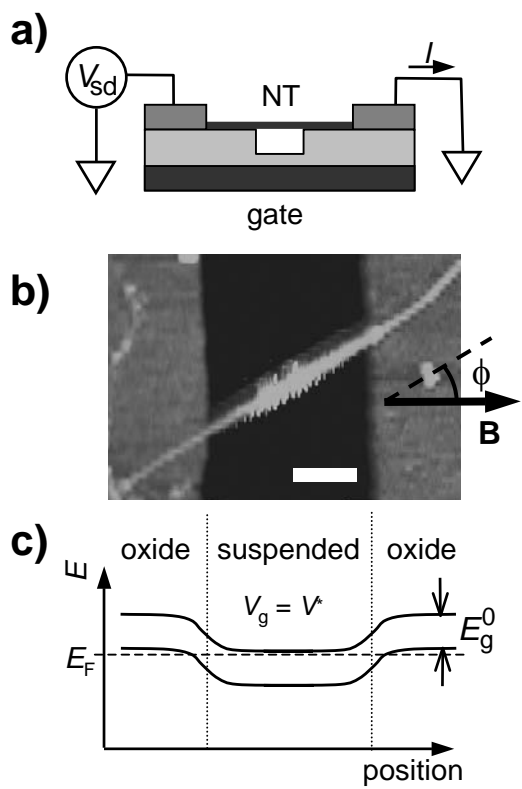

**a)** $V_{sd}$  NT  $I$  gate

**b)** $\phi$  **B**

**c)** oxide  suspended  oxide

$E$

$V_g = V^*$

$E_F$

$E_g^0$

position

**Figure 2**

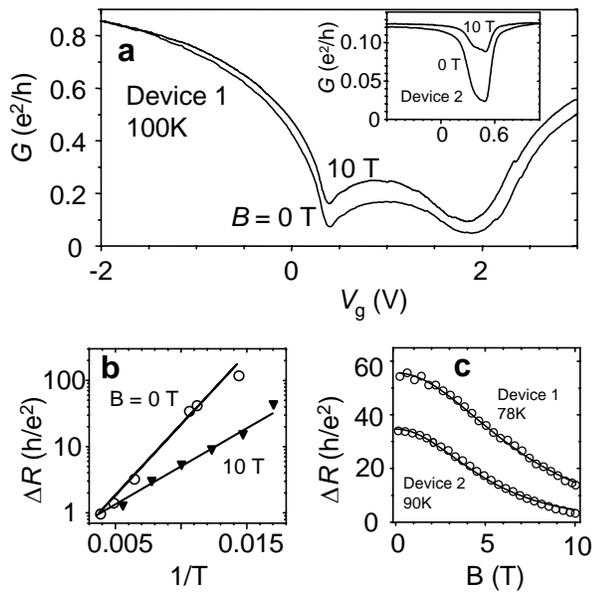

**Figure 3**

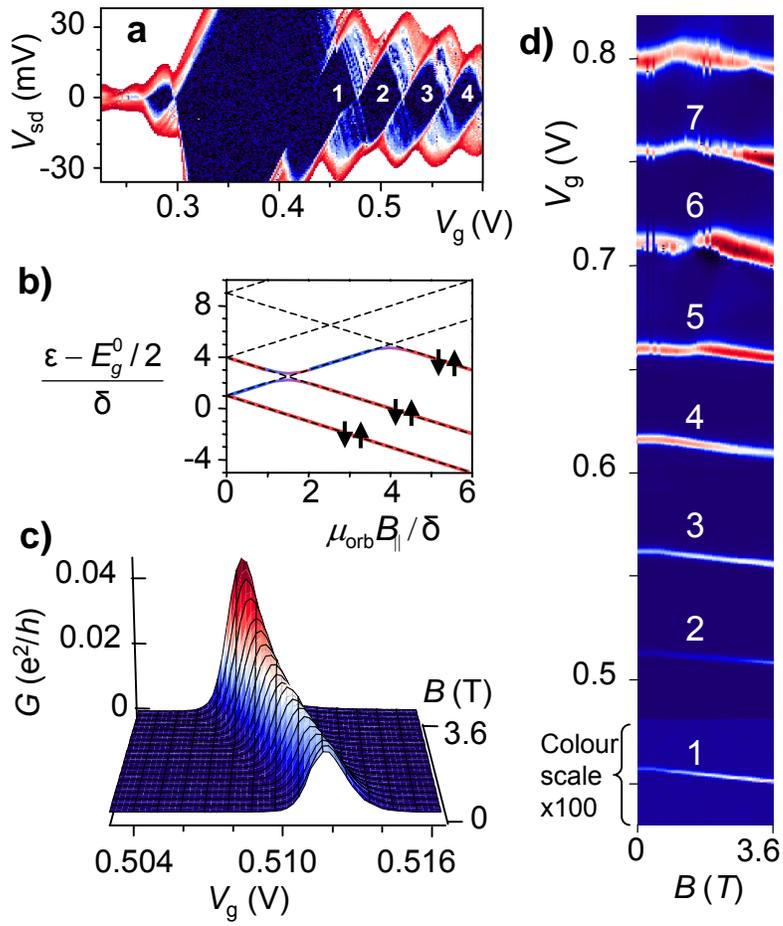

**Figure 4**